\documentclass[lettersize,journal]{IEEEtran}
\usepackage{amsmath,amsfonts}

\usepackage{algorithm}
\usepackage{algorithmic}
\usepackage{array}
\usepackage[caption=false,font=normalsize,labelfont=sf,textfont=sf]{subfig}
\usepackage{textcomp}
\usepackage{stfloats}
\usepackage{url}
\usepackage{verbatim}
\usepackage{graphicx}
\usepackage{cite}
\usepackage{multirow} 
\usepackage{xcolor}
\usepackage{booktabs}
\usepackage{amsmath}
\usepackage{amssymb}
\usepackage{soul}
\usepackage{hyperref} 

\begin{document}

\title{Automatic Detection and Analysis of Singing Mistakes for Music Pedagogy}

\author{Sumit Kumar$^{\dag}$, Suraj Jaiswal$^{\dag}$, Parampreet Singh$^{\dag}$, Vipul Arora$^{\S}$\\
        $^{\dag}$Graduate Student Member, IEEE,\quad
        $^{\S}$Senior Member, IEEE\\
        $^{\dag}$$^{\S}$Indian Institute of Technology, Kanpur\\
        $^{\S}$Department of Electrical Engineering (ESAT), KU Leuven
        

\thanks{This work is supported by IMPRINT-2C grant from DST-SERB and by Prasar Bharati.}
}

\markboth{Journal of \LaTeX\ Class Files,~Vol.~14, No.~8, August~2021}%
{Shell \MakeLowercase{\textit{et al.}}: A Sample Article Using IEEEtran.cls for IEEE Journals}


\maketitle

\begin{abstract}

The advancement of machine learning in audio analysis has opened new possibilities for technology-enhanced music education. 
This paper introduces a framework for automatic singing mistake detection in the context of music pedagogy, supported by a newly curated dataset.
The dataset comprises synchronized teacher–learner vocal recordings, with annotations marking different types of mistakes made by learners.
Using this dataset, 
we develop different deep learning models for mistake detection and benchmark them. 
To compare the efficacy of mistake
detection systems, a new evaluation methodology is proposed.
Experiments indicate that the proposed learning-based methods are
superior to rule-based methods.
A systematic study of errors
and a cross-teacher study reveal insights into music pedagogy that
can be utilised for various music applications.
This work sets out new directions of research in music pedagogy. The codes and dataset are publicly available.

\end{abstract}


\section{Introduction}

\IEEEPARstart{T}{he} proliferation of computer-assisted education methods across diverse fields has demonstrated their efficacy in enhancing learning outcomes~\cite{Kalkanoğlu_music_education_2024,kheir2023automaticpronunciationassessment,speech_accent_recognition}. Within audio-based pedagogies, such as language and music teaching, significant research efforts are underway~\cite{Kalkanoğlu_music_education_2024, piano_performance_evaluation_nature,Moura2024SoloMusicAssessment}. The field of computer-assisted language learning and pronunciation training boasts a substantial body of literature and computational resources \cite{lo2010automatic, Arora2018}.
In the case of music, although automated assessment methods have been developed for instrumental performance, such as violin and piano~\cite{piano_performance_evaluation_nature,Moura2024SoloMusicAssessment}, analogous systems for vocal pedagogy remain in their early stages of development~\cite{TISMIR_performance_analysis_review_2020}.

Indian Art Music (IAM) consists of long-established classical music traditions, including Hindustani and Carnatic music, characterized by raga and tala-based melodic and rhythmic frameworks and a strong emphasis on oral transmission. Learning is centered on imitation, memorization, and gradual refinement of pitch, rhythm, and stylistic expression through repeated listening and practice. IAM pedagogy is deeply rooted in the teacher–learner (guru–shishya) tradition, where musical knowledge is transmitted primarily through direct, in-person interaction rather than written notation~\cite{alter1997key}.
Unlike conventional classroom settings with regular instructions, music education typically involves only one or two weekly lessons, necessitating substantial independent practice between sessions~\cite{dye2016student}.
Independent practice typically relies on lesson recordings from class or on written instructions provided during the session.
But in the absence of immediate feedback, beginners may inadvertently entrench errors in pitch, rhythm, or expression during that independent practice.
Although technology offers a promising avenue to bridge this learning gap by providing immediate feedback, tracking progress, and reinforcing correct habits \cite{addessi2005experiments, webster2007computer}, systematic research into such intelligent tools specifically tailored for IAM remains limited.

This study develops an automated singing mistake detection system to support music learning. In this framework, learners attempt to replicate the teacher’s vocal performance within an \textit{imitation learning}~\cite{alter1997key} paradigm.
To enable this, we curate a dataset named M3\footnote{M3 (MADHAV Lab Mistake Detection for Music Teaching Database) available at \href{zenodo}{https://zenodo.org/records/8332078}} tailored for IAM vocal pedagogy. The dataset comprises synchronized teacher-learner singing recordings, where 
the teachers annotate instances of pitch, amplitude, pronunciation, and rhythm errors in learners' singing corresponding to the teacher's singing.

Using this dataset, we propose an automatic mistake detection framework
comprising a rule-based (RB) method,
and deep learning models
based on CNN, CRNN, and TCN architectures.
This task is formulated as an audio event detection problem \cite{mesaros2021sound, kumar2023balanced}, where the input consists of acoustic features derived from synchronized teacher and learner audio, and the target events are specific singing mistakes made by the learner. Our codes, dataset, pre-trained models, and demos are available on 
\href{https://github.com/madhavlab/2023_narottam_engine}{https://github.com/madhavlab/2023\_narottam\_engine}.

The main contributions of this work are:
\begin{itemize}
  \item The formulation of the task of automatic singing mistake detection, supported by both rule-based and deep learning approaches.  
  \item The introduction of a novel dataset for mistake detection in Indian Art Music, consisting of synchronized teacher–learner audio recordings annotated by expert teachers for pitch, amplitude, pronunciation, and rhythm-based mistakes.  
  \item The development and benchmarking of deep learning based systems for detecting melodic and amplitude errors, aimed at providing detailed corrective feedback to learners.  
  \item The development of a task-specific evaluation methodology
   and a systematic analysis of singing errors during the learning process.

\end{itemize}
\section{Related Works}

\subsection{Computers for Music Evaluation}  
Research on computer-based singing evaluation began with karaoke-style systems, comparing a subject’s performance against a reference recording and producing an overall score \cite{lal06_interspeech,Tsai2012}. Early methods emphasized pitch-based alignment, starting with direct comparisons and later using techniques such as dynamic time warping \cite{hu2002probabilistic, bozkurt2017dataset}. Subsequent works extended evaluation to other musical attributes, including rhythm and amplitude \cite{Tsai2012}, and applied classical ML methods such as SVMs and HMMs for robustness in judging pitch intervals, vibrato, and pronunciation accuracy \cite{nakano06_interspeech, Tsai2012, tsai2019automatic, miyagawa2017detection}. 

Recent approaches introduced regression and learned descriptors for instrument and voice assessments \cite{wu2016towards, vidwans2017objective, gururani2018analysis}, as well as deep networks for predicting overall performance scores \cite{pati2018assessment, huang2020spectral, li2021training, huang2020score}. Scalable assessment strategies include unsupervised clustering of large cohorts of singers by pitch, rhythm, and timbre similarity \cite{Gupta2020taslp}, while other works have addressed intelligibility in singing \cite{Sharma2020intelligibility}. Gupta \textit{et al.} \cite{gupta2022deep} provides an overview of these singing voice evaluation methods. More recently, Hsieh \textit{et al.} \cite{hsieh2025tonality} proposed a tonality-based, accompaniment-guided framework that eliminates the need for reference vocals and correlates well with human judgments.
A common tendency of these systems is to assign an overall score, overlooking details such as intonation, rhythm, or articulation, and thus offering limited feedback for music pedagogy.

\subsection{Computers for Music Education}  
Computers have been employed to support music pedagogy. Early efforts focused on real-time visualizations of learners’ pitch contours and acoustics \cite{hoppe2006development, howard1989microcomputer, welch2005real}, which proved helpful at the basic level but are often inadequate for advanced and continuous patterns, especially in Indian classical music, where auditory learning is central \cite{wilson2005looking, andrianopoulou2019aural, blake2010primacy}. Tools have also been developed for piano and keyboard pedagogy \cite{Gorbunova2019}, but the lack of sheet-music traditions in IAM limits the transferability of such approaches \cite{ranjani2019compact}.  
Beyond visualization, evaluation systems for education typically provide overall performance scores without detailed mistake feedback. Bozkurt \textit{et al.} \cite{bozkurt2018musiccritic} integrate performance analysis into an online learning platform, while several piano tutoring systems compare learner recordings to reference MIDI or scores to assess accuracy \cite{benetos2012score, ewert2016score, 7971931, ramoneda2022score}. e-learning tools further support music theory and ear training \cite{pesek2020troubadour}.

\subsection{Datasets for Music Analysis}
Saraga~\cite{Saraga}, RRD~\cite{gulati2016time}, CompMusic~\cite{serra2014creating}, and Prasarbharti Indian Music (PIM-v1)~\cite{param_PIMv1} are a few popular resources in IAM, which primarily focus on raga, tonic, and melody identification, emphasizing musical structure rather than learner performance. 
The ROD~\cite{kumar2025recognizing} dataset provides frame-wise ornamentation annotations for manually recorded IAM performances, while CoSIAN~\cite{yamamoto2022analysis} offers similar annotations for Japanese pop (J-POP) based on existing YouTube recordings. 
In the domain of automatic singing quality assessment, datasets like DAMP and NUSnQ~\cite{li2021training} contain multiple renditions of popular songs annotated with listener-based quality scores, enabling the study of reference-free evaluation and rank-ordering of singers.  
Overall, while existing datasets address structural analysis, expert ornamentation, or holistic quality assessment, they do not capture learner-specific mistakes. Our dataset instead provides synchronized teacher–learner recordings with frame-level mistake annotations, supporting fine-grained analysis for music pedagogy.

Most prior works in music education and evaluation have focused on assigning holistic performance scores, offering little support for detailed mistake detection or targeted feedback. Existing systems are also developed mainly for Western music with discrete note structures, leaving IAM
largely unexplored. To address this, we introduce a dataset and framework for frame-wise mistake detection in IAM, built on synchronized teacher–learner recordings with fine-grained annotations.

\section{Dataset}
\subsection{Recording Setup}
The recording setup for features a teacher singing with an in‑ear drone instrument (\textit{t\=anpur\=a}) to provide melodic reference, and a percussion playback (tabla) for rhythmic support.
Learner recordings follow a similar process, with the addition of the teacher’s pre-recorded voice, pitch-shifted to match the learner’s tonic.
Thus, the in‑ear mix contains the teacher's voice, drone, and percussion, while only the learner’s vocal is recorded.
Since frame-wise mistake analysis requires precise synchronization between teacher and learner renditions, we ensure time alignment during playback and recording. Playback latency in our environment is stable and corrected via a simple correlation-based calibration; for more general settings with variable latency, robust alignment methods such as \cite{10096874} may be required.  
All recordings are made with an Audio-Technica AT2020 cardioid microphone and a laptop with a 12th Gen Intel Core i5-12500H processor. Audio capture is handled in Audacity, with audio stored as mono WAV files at 44.1 kHz, 32-bit precision. Informed consent was obtained from all human participants involved in the study, and the study protocol was approved by the institutional ethics committee.  

\begin{table}[t]
\caption{M3 Dataset Statistics}
\centering
\begin{tabular}{|l|l|l|}
\hline
\textbf{}             & \textbf{Teacher 1} & \textbf{Teacher 2} \\ \hline
No. of Teacher Files  & 64                 & 91                 \\ \hline
Total duration (in s) & 1125.5           & 5483.7           \\ \hline
No. of Learners Files & 259                & 242                \\ \hline
Total duration (in s) & 4623.7           & 14956.4          \\ \hline
\end{tabular}
\label{table:data_description}
\end{table}

\begin{figure}[t]
\centering
\includegraphics[width=0.87\columnwidth]
{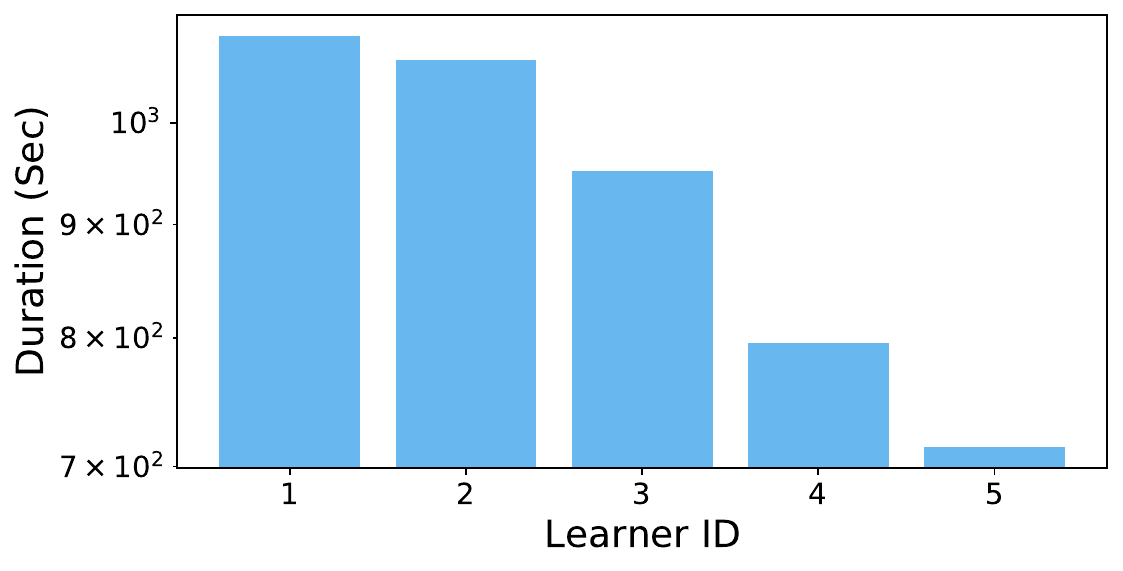}
\caption{\label{fig:learner_dist_001}{Duration-wise distribution of recordings across learners associated with Teacher 1. The Learner ID (x-axis) represents an anonymized identifier assigned to each learner.}}
\end{figure}

\begin{figure}[t]
\centering
\includegraphics[width=0.87\columnwidth]{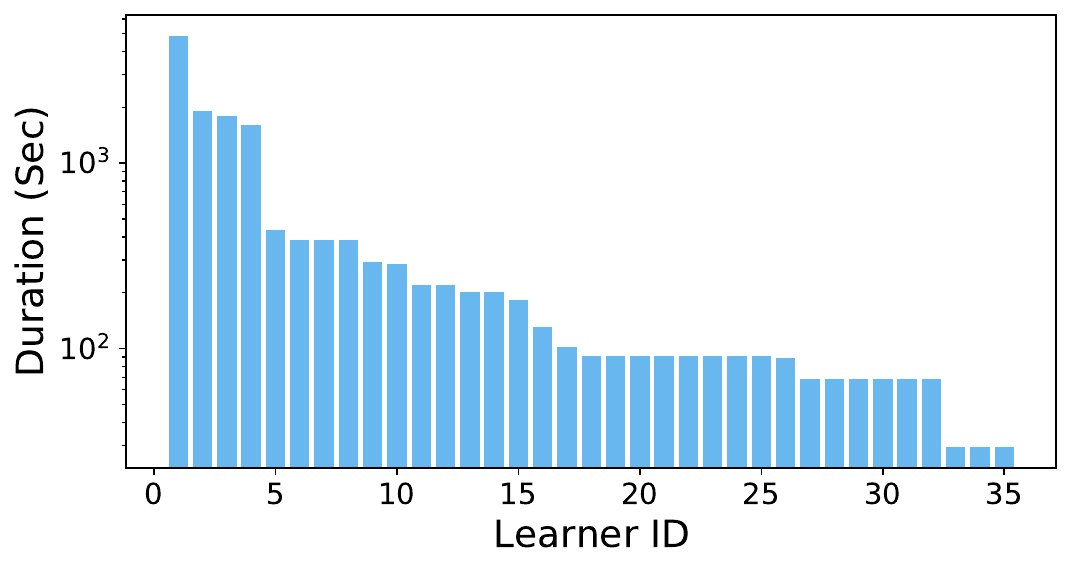}
\caption{\label{fig:learner_dist_002}{Duration-wise distribution of recordings across learners associated with Teacher 2. The Learner ID (x-axis) represents an anonymized identifier assigned to each learner.}}
\end{figure}

\begin{figure}[t]
\centering
\includegraphics[width=0.87\columnwidth]{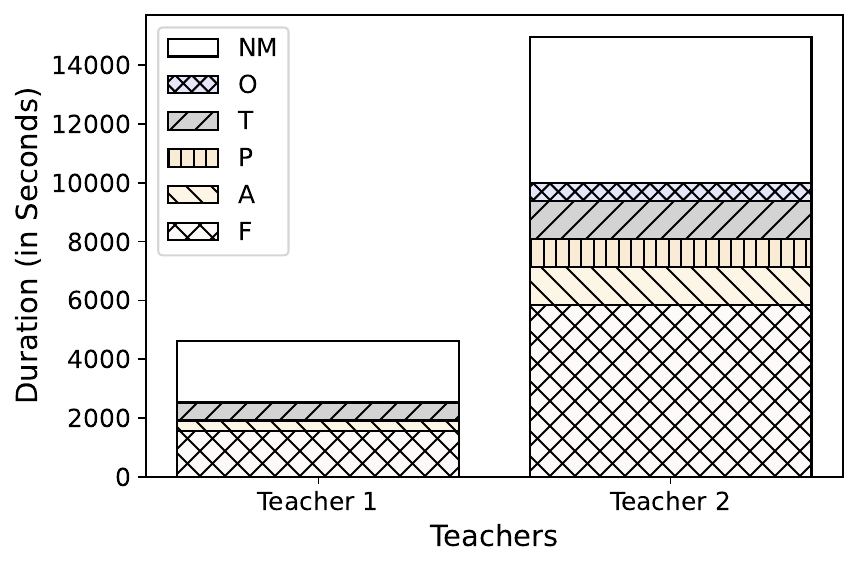}
\caption{\label{fig:mistake_dist}{Ground truth mistake distribution for all learners of each teacher where F: frequency mistakes, A: amplitude mistakes, P: pronunciation mistakes, T: timing mistakes, O: other mistakes, NM: no mistake}}
\end{figure}

\subsection{Dataset and Metadata}
The dataset comprises recordings from two Indian classical vocal teachers and their respective, non-overlapping sets of learners, as summarized in Table~\ref{table:data_description}. Each teacher recorded a sequence of lessons, progressing from simple beginner exercises to more complex melodic and rhythmic structures. 
Each teacher's audio file corresponds to a single lesson. 

Learners followed the teacher’s sequence for recording, practicing each lesson multiple times until it felt stable, then making a single recording before moving to the next lesson.
Learners practiced from the teacher’s reference audio and a written transcription of the lesson.
The average pace for learners was 1–2 lessons per day over 2–3 months. No real-time feedback was provided during takes, deliberately mimicking unsupervised practice. This setup preserves increasing complexity, supports fair comparison across learners, and reflects real pedagogy where mastery of earlier material underpins later exercises. Learners record the lessons in the same prescribed order, simulating a progression of training without the real-time presence of a teacher. All learners are beginners with no formal music training, though some have casual or informal singing experience. This ensures that the data realistically captures the kinds of mistakes typical at the beginner level.  

Teacher metadata includes tonic, tempo in Beats Per Minute (BPM), \textit{t\=ala} name, and lesson identifier. Learner metadata includes the same fields with additional entries for
the reference teacher lesson, onset of the first beat (\textit{sama}), and the corresponding note sequence, transcribed in the \textit{Bh\=atkhand\=e} system \cite{Ome}. The number of recordings per learner is variable and shown in Figs.~\ref{fig:learner_dist_001} and \ref{fig:learner_dist_002}. 
Each learner’s audio is accompanied by a text annotation file containing time-stamped mistakes annotated by the respective teacher. These annotations are used as ground-truth labels for our analysis, where frames corresponding to annotated mistake intervals are assigned a label of 1 and all remaining frames are assigned 0, resulting in a multi-label learning setting.
The ground truth mistake distribution across recordings is illustrated in Fig.~\ref{fig:mistake_dist}.  

The dataset represents beginner-level, imitation-based pedagogical exercises in IAM, where learning emphasizes accurate reproduction of the teacher’s rendition in pitch, rhythm, and clarity, deviation from which is annotated as a mistake. 
The dataset does not aim to cover advanced improvisational performance, concert renditions, or stylistically flexible interpretations in IAM.

\subsection{Mistake Annotation}
Two expert teachers provide independent annotations of learner mistakes for their respective student recordings.
For each learner–teacher pair, the teacher assesses both renditions and marks start time, end time, and a mistake category label:  
\begin{itemize}
\item \textbf{Frequency mistake:} Perceptible deviation of the learner’s pitch from
the target note of the teacher.
    \item \textbf{Amplitude mistake:} Audible discrepancy in loudness or energy
between teacher and learner.
    \item \textbf{Pronunciation mistake:} Incorrect articulation of the syllable (e.g., \textit{Sa} pronounced as \textit{Pa}). 
    \item \textbf{Timing mistake} Loss of synchrony with the accompanying rhythmic cycle.

\end{itemize}
Less significant errors are marked as \textbf{Others}, followed by a description.
Mistake annotation in vocal music is inherently subjective, since deviations from the reference may reflect stylistic or expressive differences rather than pedagogical errors. Annotators therefore focus on aspects considered most relevant for beginners, such as pitch stability, timing accuracy, and clarity of pronunciation, while minor variations in ornamentation are often disregarded to avoid learner's cognitive overload. Consequently, annotations are teacher-specific, and inter-teacher consistency is not assumed.  

To ensure high-quality annotations, we adopt a two-tier verification process inspired by \cite{kumar2025recognizing}.
Initially, each teacher provides annotations for their corresponding learner recordings. We then train preliminary deep models using these first-round annotations and compare their predictions with the original teacher labels. Disagreements between model predictions and teacher annotations are flagged and sent back to the annotating teacher for re-evaluation. Each learner recording undergoes two rounds of such verification, with teachers refining their annotations based on systematic disagreements identified by the automated analysis. This iterative human-in-the-loop approach yields a more consistent and refined annotation set while preserving teacher-specific assessment criteria.
The design of the dataset captures realistic learning trajectories of beginners in IAM

Although the dataset includes four categories of vocal mistakes, pitch and amplitude-related deviations are the most consistently observed and are fundamental to vocal performance. Timing-related errors require dedicated modeling schemes, while pronunciation errors are sparse due to the use of written lesson transcriptions during practice. Therefore, for this study, we center our focus only on pitch and amplitude-related mistakes.

\section{Methodology}

\subsection{Problem Setup}
Let the dataset be defined as
\[
\mathcal{D} = \Big\{ \big(x^t_j,\; x^\ell_{jk},\; \mathbf{Y}_{jk}\big) 
\;\Big|\; 
j \in \{1,\dots,N\},\; k \in \{1,\dots,M_j\} \Big\},
\]
where $x^t_j$ denotes the $j^{\text{th}}$ teacher audio, 
$x^\ell_{jk}$ denotes the $k^{\text{th}}$ learner rendition corresponding to teacher audio $j$, 
$N$ is the total number of teacher audio, 
and $M_j$ is the number of learner renditions associated with teacher $j$. 
$\mathbf{Y}_{jk}$ denotes the frame-level ground-truth annotations for the pair $(x^t_j, x^\ell_{jk})$, and is given by:

\[
\mathbf{Y}_{jk} = \left\{ \big(s_{jkm},\, e_{jkm},\, c_{jkm}\big) \right\}_{m=1}^{L_{jk}},
\]
where $L_{jk}$ denotes the total number of mistakes annotated for the pair $(x^t_j, x^\ell_{jk})$. 
Here, $s_{jkm}$ and $e_{jkm}$ represent the onset and offset times (in seconds) of the $m^{\text{th}}$ mistake, 
and $c_{jkm} \in \{\text{F}, \text{A}\}$ indicates its class: Frequency (F) or Amplitude (A).

Separate acoustic feature sets are extracted for frequency and amplitude mistakes as discussed in the following section. For brevity, subscripts $j, k, m$ are omitted henceforth.

\subsection{Feature Representation}
\label{section: features}
\subsubsection{Pitch Features}
For frequency-based mistake detection, pitch contour $p[n]$ is extracted using Praat-Parselmouth python package\footnote{https://pypi.org/project/praat-parselmouth/} with a 60 ms analysis window and a 10 ms hop, where $n$ denotes the frame index of the pitch contour.
Since octave errors (e.g., the correct note in a different octave) are not treated as mistakes in musicology, we normalize the pitch contour $p[n]$ by the tonic $s$ of the corresponding audio $x$
to obtain an octave‑invariant representation.
To avoid ambiguity between voiced and unvoiced frames, silent frames are explicitly assigned a sentinel value outside the valid pitch range:
\begin{equation}
\label{eq: tonic_norm}
\hat{p}[n] = 
\begin{cases} 
\log_2\left(\frac{p[n]}{s}\right) \% 1 & \text{if } p[n] > 0 \\ 
-1 & \text{if } p[n] = 0 
\end{cases}
\end{equation}
Here, voiced frames satisfy $\hat{p}[n] \in [0,1)$, while unvoiced frames are mapped to $-1$, ensuring clear separation between silence and valid pitch values.
We further experiment with
chromagram features
extracted using a 2048-point STFT with a 46-ms Hanning window and a 23-ms hop size.

\subsubsection{Amplitude Features}
For amplitude-based mistakes, we compute short-time RMS (root mean square) energy:
\begin{equation}
\label{rmse}
    a[n] = \sqrt{\frac{1}{N} \sum_{m=1}^{N} x[n-m]^2},
\end{equation}
followed by log compression:
$
\hat{a}[n] = \log_2(a[n]).
$
where $N$ denotes the length of $x[n]$.

Amplitude mistakes are relatively rare in real performances, leading to class imbalance while training deep-learning models. 
To address this, we augment learner audio by introducing artificial amplitude errors. We manually modify the energy of stable note segments obtained via the segmentation procedure as described next.

\paragraph{Data Augmentation}
We perform data augmentation for amplitude mistakes at the note level, which requires segmenting notes from the learner audio as a first step. The pitch contour is extracted, smoothed, and mapped to the MIDI scale, from which note boundaries are identified and candidate notes obtained. Each segment is then analyzed for energy and pitch stability to classify it as silence, transition, or static. Finally, amplitude-related mistakes are simulated by perturbing the energy of static notes through random scaling, generating attenuated and amplified variants that enrich the dataset and mitigate class imbalance during training. This procedure is explained in detail as follows.

Given a pitch contour $p[n] \in \mathbb{R}^+$ sampled at regular intervals $\Delta t$ seconds, the corresponding time at frame $n$ is $t_n = n \cdot \Delta t$. To account for unvoiced or silent frames, we replace $p[n] = 0$ with a small positive constant $\epsilon = 10^{-6}$. The resulting sequence is smoothed using a median filter of window size $w_m$ (e.g., $w_m = 20$).
This smoothed pitch contour $p[n]$ is then mapped to the MIDI scale as:

\begin{equation}
\label{midi}
m[n] = 69 + 12 \cdot \log_2\left(\frac{p[n]}{440}\right)
\end{equation}

To segment notes, we traverse the MIDI sequence $m[n]$ from left to right. A new note boundary is defined at frame $n$ if the pitch deviates from the reference pitch by more than a defined threshold $\delta$ (e.g., $\delta = 0.5$ semitones):
\begin{equation}
    \label{note_segments}
    |m[n] - m[n_0]| > \delta
\end{equation}

Here, $n_0$ is the reference frame. If the duration of the segment from $t_{n_0}$ to $t_n$ exceeds a minimum length $T_{\min}$ (e.g., $0.5$ seconds), it is retained as a candidate note. This process yields a list of note segments:
\begin{equation}
    \label{S}
    \mathcal{S} = \left\{ (t_k^{\text{on}}, t_k^{\text{off}}) \right\}_{k=1}^{K}
\end{equation}

To further categorize each segment, we analyze both pitch stability and energy content. Let $x[n]$ denote the raw audio signal sampled at $F_s$ Hz. The average of root mean square energy, $\bar{E}_k$ is computed over non-overlapping frames as given in Eq.~\ref{rmse}.

For pitch analysis, let $\mathbf{m}_k$ be the MIDI values in the $k^{\text{th}}$ segment. To ensure robustness, we discard the first and last $w_p$ frames (e.g., $w_p = 15$), and compute the standard deviation:

\begin{equation}
    \label{sigma_k}
    \sigma_k = \text{std}\left(\mathbf{m}_k[w_p : -w_p]\right)
\end{equation}

Each segment is classified into one of three categories based on the following rules:

\begin{equation}
   \label{label_k}
\text{label}_k =
\begin{cases}
\text{silence}, & \text{if } \bar{E}_k < \tau_E \\
\text{transition}, & \text{if } \sigma_k > \tau_{\sigma} \\
\text{static}, & \text{otherwise}
\end{cases} 
\end{equation}

where $\tau_E$ = 0.01 and $\tau_{\sigma}$ = 10.0 are empirically chosen thresholds for energy and pitch variance, respectively.

To augment the dataset with amplitude-related mistakes, we perturb the energy of stable (static) note segments by scaling with a random factor, $ r_k \sim \mathcal{U}[0,2]$, where $r_k < 1$ attenuates and $r_k > 1$ amplifies the segment. Both the amplified and attenuated segments count as augmented amplitude error. This procedure is summarized in Algorithm~\ref{algo: note}.

\begin{figure*}[t]

\centering
\includegraphics[width=2\columnwidth]{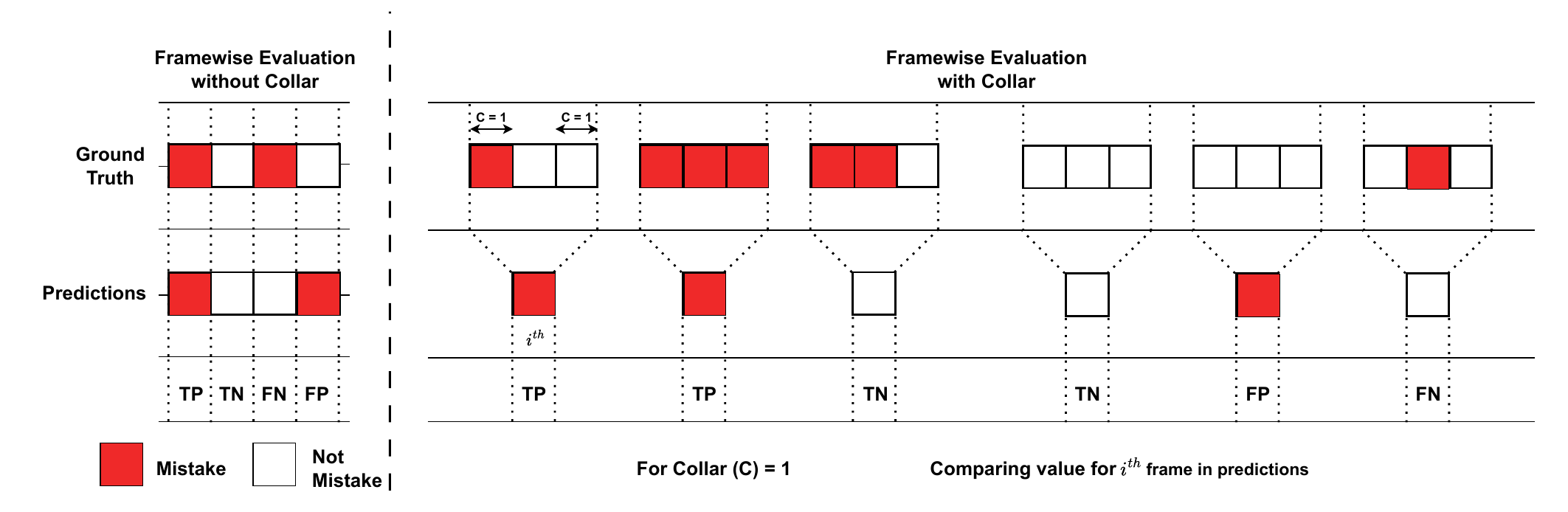}
\caption{\label{fig:evalatuion}{Illustration of collar-based frame-wise evaluation. Left: naive frame-wise evaluation without collars. Right: with a collar of $c=1$ frame, ground-truth mistake frames are dilated by one frame on both sides (gray arrows). For each predicted mistake frame (bottom row), we assign: True Positive (TP) if it overlaps any dilated ground-truth mistake frame, False Positive (FP) if it does not overlap any, and False Negative (FN) for a ground-truth mistake frame that is not predicted as a mistake even after dilation. True Negatives (TN) may be omitted as they are not used in metric computation.}}
\end{figure*}

\begin{algorithm}[t]
\caption{Note Segmentation and Augmentation}
\label{algo: note}

\begin{algorithmic}[1]
\REQUIRE Pitch contour $p[n]$, audio signal $x[n]$, thresholds $\epsilon, \delta, T_{\min}, \tau_E, \tau_{\sigma}$
\STATE Smooth $p[n]$ (replace zeros with $\epsilon$, apply median filter) and convert to MIDI $m[n]$ using Eq.~\eqref{midi}.
\STATE Segment notes by traversing $m[n]$; create $\mathcal{S}$ using Eq.~\eqref{note_segments}--\eqref{S}.
\FOR{each segment $k \in \mathcal{S}$}
    \STATE Compute energy $\bar{E}_k$ and pitch std.\ $\sigma_k$ using Eq.~\eqref{rmse}--\eqref{sigma_k}.
    \STATE Assign $\text{label}_k$ as in Eq.~\eqref{label_k}.
\ENDFOR
\STATE Identify $\mathcal{K}_{\mathrm{static}} \leftarrow \{k \in \mathcal{S} \mid \text{label}_k = \text{static}\}$.
\STATE Randomly select $M$ notes $\mathcal{K}_M \subset \mathcal{K}_{\mathrm{static}}$.
\FOR{each $k \in \mathcal{K}_M$}
    \STATE Sample $r_k \sim \mathcal{U}(0.2,0.6) \cup \mathcal{U}(1.2,1.8)$
    \STATE Scale: $x_k^{\mathrm{aug}}[n] = r_k \cdot x[n], \; n \in [t_k^{\mathrm{on}}, t_k^{\mathrm{off}})$
\ENDFOR
\RETURN $x_k^{\mathrm{aug}}[n]$
\end{algorithmic}
\end{algorithm}

\subsection{Models}

\subsubsection{Baseline} \label{sec:Baseline}
As a rule-based (RB) baseline, we compare the teacher and learner contours at each frame. A frequency error is flagged if  
$
|\hat{p}^{t}[n] - \hat{p}^{\ell}[n]| > \beta
$,
with $\beta$ being the frequency threshold. Similarly, an amplitude error is flagged if
$
|\hat{a}^{t}[n] - \hat{a}^{\ell}[n]| > \gamma
$,
with $\gamma$ being the amplitude threshold. Both thresholds are optimized via grid search on the training set.

\subsubsection{Deep Learning-Based models}
For deep learning-based models, the inputs consist of stacked teacher–learner pitch contours $(\hat{p}^t[n], \hat{p}^\ell[n])$ and stacked chromagram features for frequency mistakes; amplitude contours $(\hat{a}^t[n], \hat{a}^\ell[n])$ for amplitude mistakes.
All models share the same architecture and training setup, except for the type of convolution used. For contour-based inputs (pitch or amplitude sequences), we apply 1D convolutions, while for chromagram features, we use 2D convolutions.
The mistake detection task is formulated as binary classification, where a time-distributed dense layer with sigmoid activation produces frame-level outputs. The models are trained using the weighted binary cross-entropy loss defined as:

\[
\mathcal{L} = - \frac{1}{T} \sum_{n=1}^{T} \left[ w_1 y[n] \log \hat{y}[n] + w_0 (1 - y[n]) \log (1 - \hat{y}[n]) \right]
\]

where $y[n] \in \{0,1\}$ is the ground-truth label, $\hat{y}[n] \in [0,1]$ is the predicted probability, and $w_0, w_1$ are class weights computed as the inverse of their respective class frequencies.
All models are trained with the Adam optimizer with learning rate = 0.001, batch size 32, and early stopping based on validation loss with a patience of 10. 

\paragraph{CNN Model}
A 1D-CNN model with five convolutional layers (filter sizes $[8,16,32,32,8]$, kernel size 3), each followed by batch normalization, ReLU, and Max Pooling.

\paragraph{CRNN Model}
A CRNN model with three convolutional layers (filter sizes $[8,16,8]$, kernel size $3$), each with batch normalization, ReLU activation, and max pooling. The convolutional output is fed to a bidirectional GRU (8 units).

\paragraph{Temporal Convolutional Network (TCN)}
Finally, we experiment with a TCN~\cite{lea2017temporalTCN} employing an encoder–decoder architecture with dilated convolutions to capture long-range dependencies. The encoder (having 3 convolutional layers [8, 16, 32])  progressively downsamples via max pooling, while the decoder (consisting of 3 convolutional layers [32, 16, 8]) upsamples back to frame resolution. Each convolutional layer uses ReLU activation and a kernel of size 3.

\subsection{Evaluation Metrics}

The performance of singing mistake detection methods can be evaluated using standard detection metrics such as precision and recall at a frame level. However, the nature of the task poses inherent difficulties for such idealized frame-level matching. Since it is exceedingly laborious and practically infeasible to mark the exact onset and offset of a mistake, it is not possible to label each frame of the audio representation as correct or incorrect precisely. This limitation has been extensively addressed in the audio event detection literature~\cite{mesaros2021sound}, where evaluation is based on event-based metrics. 
Following this approach, we adopt a collar-based annotation scheme with soft temporal boundaries for ground-truth mistakes, 
where each annotated mistake is softened by symmetrically dilating the interval on both sides by a duration $T_c$.
So, we get a collar of $c$ frames given by:
\begin{equation}
c = \frac{T_c}{\Delta t}.
\end{equation}

\begin{figure*}[t]
\centering
\subfloat[{\small For Teacher 1}]{\includegraphics[width=0.87\columnwidth]{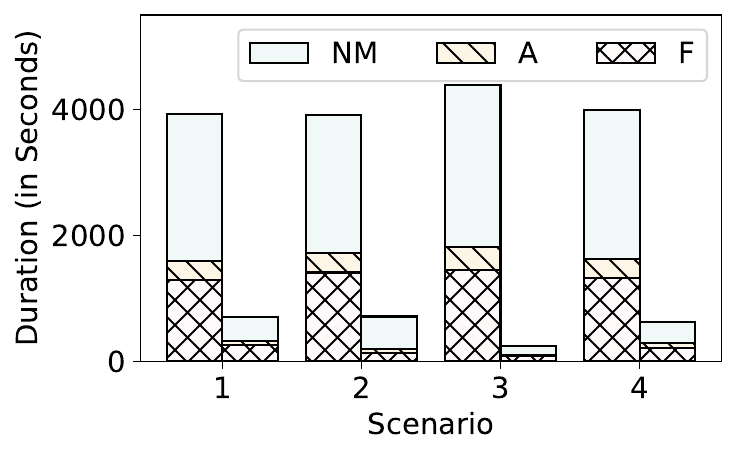}%
\label{fig:scenario_dist_001}}
\hfill
\subfloat[{\small For Teacher 2}]{\includegraphics[width=0.87\columnwidth]{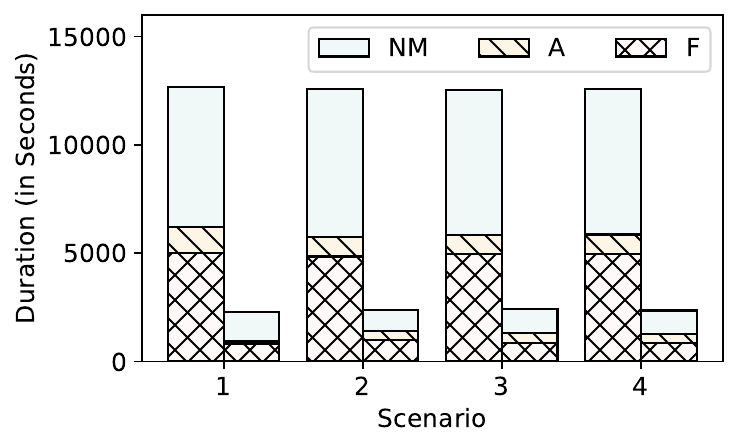}%
\label{fig:scenario_dist_002}}
\caption{Ground truth class-distribution for all four split scenarios. Within each scenario, the bar on the left corresponds to the training set and the bar on the right corresponds to the test set. F: frequency mistakes, A: amplitude mistakes, NM: no mistake.}
\label{fig:scenario_dist}
\end{figure*}

For evaluation, each ground-truth mistake frame is first marked, and then its neighborhood is dilated by $c$ frames on both sides. We then compare the binary prediction at each frame to this dilated ground-truth to define:
\begin{itemize}
    \item    \textbf{True Positive}: a predicted mistake frame that falls within the dilated ground-truth mistake region (i.e., has non-zero overlap with a collared ground-truth mistake).
    \item    \textbf{False Positive}: a predicted mistake frame that does not fall within any dilated ground-truth mistake region.
    \item     \textbf{False Negative}: a ground-truth mistake frame whose prediction is not a mistake, even after dilation of the ground-truth region.
    \item   \textbf{True Negatives}: a predicted no-mistake frame that does not overlap with collared ground-truth mistake region.
    
\end{itemize}

The overall evaluation methodology is illustrated in Fig.~\ref{fig:evalatuion}.
We also report event-based metrics \cite{mesaros2021sound}, where predicted mistake events are matched to annotated mistake events based on temporal overlap. Unlike frame-based evaluation, this metric assesses detection performance at the level of entire events. It can be more sensitive to over-segmentation, which is addressed using dilated ground-truth boundaries with a collar-based approach.

\begin{table*}[t]
\caption{\label{table:results_frame_f_t1}Teacher-1 F-Mistakes. 
Results under Without Collar and With Collar correspond to frame-based metrics,
while the Event-Based section reports event-based metrics. Results showing performance of the rule-based model (RB), models using pitch contour features (CNN, CRNN (P), TCN (P)), and models using chromagram features (CRNN, TCN). 
}
\centering
\begin{tabular}{|c|c|ccc|ccc|ccc|ccc|}
\hline
 & \multicolumn{1}{c|}{\multirow{2}{*}{\textbf{Model}}} & \multicolumn{3}{c|}{\textbf{Scenario-1}} & \multicolumn{3}{c|}{\textbf{Scenario-2}} & \multicolumn{3}{c|}{\textbf{Scenario-3}} & \multicolumn{3}{c|}{\textbf{Scenario-4}} \\
\cline{3-14}
 &  & \textbf{F1} & \textbf{P} & \textbf{R} & \textbf{F1} & \textbf{P} & \textbf{R} & \textbf{F1} & \textbf{P} & \textbf{R} & \textbf{F1} & \textbf{P} & \textbf{R} \\
\hline
\multirow{6}{*}{\rotatebox[origin=c]{90}{\textbf{Without Collar}}}
 & \textbf{RB}        & 61.69 & 75.52 & 52.14 & 43.81 & 84.44 & 29.58 & 62.19 & 76.22 & 52.53 & 80.47 & 92.42 & 71.26 \\ 
 & \textbf{CNN (P)}   & 71.10 & 71.00 & 71.20 & 68.28 & 79.36 & 59.92 & 73.23 & 81.10 & 66.76 & 77.84 & 74.45 & 81.64 \\ 
 & \textbf{CRNN}      & 71.35 & 92.61 & 58.02 & 52.16 & 48.79 & 56.02 & 67.81 & 97.46 & 52.00 & 77.66 & 88.20 & 69.37 \\ 
 & \textbf{TCN}       & 70.06 & 86.00 & 59.11 & 54.32 & 86.32 & 39.63 & 67.79 & 92.79 & 53.41 & 72.51 & 93.18 & 59.34 \\ 
 & \textbf{CRNN (P)}  & 74.94 & 79.67 & 70.75 & 69.36 & 93.81 & 55.02 & 76.76 & 94.35 & 64.70 & 86.71 & 96.35 & 78.83 \\ 
 & \textbf{TCN (P)}   & \textbf{79.53} & 95.24 & 68.27 & \textbf{75.66} & 93.80 & 63.39 & \textbf{78.81} & 97.62 & 66.08  & \textbf{86.86} & 99.40 & 77.13  \\ 
\hline
\multirow{6}{*}{\rotatebox[origin=c]{90}{\textbf{With Collar}}}
 & \textbf{RB}        & 64.47 & 78.35 & 54.76 & 46.78 & 88.51 & 31.79 & 65.31 & 78.87 & 55.74 & 82.56 & 94.00 & 73.60 \\
 & \textbf{CNN (P)}   & 74.10 & 73.56 & 74.65 & 72.36 & 83.49 & 63.85 & 76.47 & 83.52 & 70.52 & 80.50 & 76.73 & 84.67 \\
 & \textbf{CRNN}      & 77.20 & 96.26 & 64.44 & 62.03 & 59.74 & 64.50 & 74.46 & 98.76 & 59.76 & 82.47 & 89.78 & 76.26 \\
 & \textbf{TCN}       & 76.30 & 89.22 & 66.64 & 63.29 & 91.91 & 48.26 & 74.29 & 95.33 & 60.86 & 78.13 & 94.44 & 66.63 \\
 & \textbf{CRNN (P)}  & 78.08 & 82.10 & 74.44 & 72.88 & 99.26 & 57.58 & 81.86 & 98.02 & 70.28 & 89.47 & 99.75 & 81.11 \\
 & \textbf{TCN (P)}   & \textbf{82.47} & 96.54 & 71.98 & \textbf{79.11} & 95.32 & 67.62 & \textbf{80.35} & 92.72 & 70.89 & \textbf{86.48} & 98.63 & 77.00 \\
\hline
\multirow{6}{*}{\rotatebox[origin=c]{90}{\textbf{Event Based}}}
 & \textbf{RB}        & 9.43 & 11.23 & 8.13 & 9.13 & 10.00 & 8.41 & 16.87 & 18.92 & 15.22 & 26.67 & 27.59 & 25.81  \\
 & \textbf{CNN (P)}   & 13.31 & 9.77 & 35.78 & 23.26 & 18.51 & 55.00 & 21.98 & 17.16 & 41.28 & 29.15 & 24.57 & 46.49  \\
 & \textbf{CRNN}      & 44.11 & 48.62 & 40.37 & 37.27 & 34.95 & 39.92 & 39.18 & 37.21 & 41.38 & 53.06 & 50.44 & 55.98  \\
 & \textbf{TCN}       & 44.64 & 52.98 & 38.57 & 37.17 & 29.53 & 50.13 & 48.87 & 42.82 & 56.90 & 56.72 & 44.55 & 78.03  \\
 & \textbf{CRNN (P)}  & 37.30 & 33.38 & 48.74 & 38.33 & 35.83 & 44.11 & 41.78 & 37.11 & 53.49 & 61.59 & 59.29 & 67.84 \\
 & \textbf{TCN (P)}   & \textbf{52.93} & 51.40 & 57.35 & \textbf{39.12} & 36.86 & 44.12 & \textbf{46.70} & 45.74 & 48.84 & \textbf{67.51} & 66.80 & 69.46  \\
\hline
\end{tabular}
\end{table*}

\begin{table*}[t]
\caption{\label{table:results1c_f}{Teacher-2 F-Mistakes}}
\centering
\begin{tabular}{|c|c|ccc|ccc|ccc|ccc|}
\hline
 & \multicolumn{1}{c|}{\multirow{2}{*}{\textbf{Model}}} & \multicolumn{3}{c|}{\textbf{Scenario-1}} & \multicolumn{3}{c|}{\textbf{Scenario-2}} & \multicolumn{3}{c|}{\textbf{Scenario-3}} & \multicolumn{3}{c|}{\textbf{Scenario-4}} \\
\cline{3-14}
 &  & \textbf{F1} & \textbf{P} & \textbf{R} & \textbf{F1} & \textbf{P} & \textbf{R} & \textbf{F1} & \textbf{P} & \textbf{R} & \textbf{F1} & \textbf{P} & \textbf{R} \\
\hline
\multirow{6}{*}{\rotatebox[origin=c]{90}{\textbf{Without Collar}}}
 & \textbf{RB}        & 65.14 & 85.00 & 52.81 & 70.56 & 87.24 & 59.24 & 67.00 & 83.90 & 55.77 & 63.70 & 84.19 & 51.23 \\ 
 & \textbf{CNN (P)}   & 72.42 & 64.00 & 83.42 & 80.24 & 72.89 & 89.27 & 75.96 & 67.57 & 88.73 & 74.47 & 64.94 & 87.29 \\ 
 & \textbf{CRNN}      & 72.71 & 63.06 & 85.86 & 78.14 & 72.35 & 84.96 & 79.00 & 82.45 & 75.82 & 72.72 & 65.87 & 81.16 \\ 
 & \textbf{TCN}       & 73.75 & 87.26 & 63.87 & 78.59 & 86.99 & 71.68 & 78.37 & 91.48 & 68.55 & 76.30 & 85.90 & 68.64 \\ 
 & \textbf{CRNN (P)}  & 84.34 & 93.63 & 76.72 & 80.27 & 91.79 & 71.33 & 79.85 & 83.14 & 76.81 & 84.59 & 89.91 & 79.87 \\ 
 & \textbf{TCN (P)}   & \textbf{87.15} & 97.17 & 79.00 & \textbf{90.60} & 95.05 & 86.54  & \textbf{89.54} & 99.11 & 81.66 & \textbf{86.91} & 92.02 & 82.18 \\ 
\hline
\multirow{6}{*}{\rotatebox[origin=c]{90}{\textbf{With Collar}}}
 & \textbf{RB}        & 66.76 & 86.24 & 54.46 & 71.69 & 88.35 & 60.32 & 68.21 & 85.00 & 56.96 & 65.00 & 85.39 & 52.48 \\ 
 & \textbf{CNN (P)}   & 74.41 & 65.91 & 85.43 & 81.77 & 74.36 & 90.83 & 77.54 & 69.10 & 88.32 & 76.14 & 66.60 & 88.88 \\ 
 & \textbf{CRNN}      & 76.99 & 67.90 & 88.89 & 81.62 & 75.76 & 88.47 & 82.29 & 85.07 & 79.68 & 76.58 & 70.09 & 84.39 \\
 & \textbf{TCN}       & 77.47 & 89.48 & 68.31 & 81.64 & 89.07 & 75.35 & 81.27 & 93.19 & 72.04 & 79.80 & 88.26 & 72.81  \\
 & \textbf{CRNN (P)}  & 86.13 & 94.47 & 79.15 & 81.83 & 93.20 & 72.93 & 81.37 & 84.92 & 78.10 & 86.19 & 90.78 & 82.05 \\ 
 & \textbf{TCN (P)}   & \textbf{87.14} & 98.44 & 78.16 & \textbf{88.35} & 99.39 & 79.51 & \textbf{87.94} & 99.68 & 78.67 & \textbf{88.14} & 98.72 & 79.61 \\
\hline
\multirow{6}{*}{\rotatebox[origin=c]{90}{\textbf{Event Based}}}
 & \textbf{RB}        & 13.96 & 18.53 & 11.20 & 15.34 & 11.91 & 21.54 & 12.94 & 10.59 & 16.62 & 10.36 & 7.10 & 19.15 \\
 & \textbf{CNN (P)}   & 16.64 & 22.47 & 13.21 & 25.61 & 34.18 & 20.47 & 14.24 & 22.70 & 10.37 & 21.72 & 28.94 & 17.38  \\
 & \textbf{CRNN}      & 34.81 & 34.64 & 34.98 & 40.08 & 40.11 & 40.06 & 45.78 & 44.41 & 46.39 & 36.42 & 36.01 & 36.84 \\
 & \textbf{TCN}       & 33.74 & 30.94 & 37.10 & 49.04 & 45.31 & 53.43 & 46.32 & 43.44 & 49.60 & 41.24 & 39.59 & 43.04 \\
 & \textbf{CRNN (P)}  & 46.35 & 43.35 & 53.95 & 50.91 & 53.80 & 48.31 & 44.02 & 48.67 & 40.18 & 45.18 & 42.44 & 51.10 \\
 & \textbf{TCN (P)}   & \textbf{58.80} & 57.64 & 61.19 & \textbf{65.33} & 63.90 & 68.60 & \textbf{63.88} & 63.68 & 63.85 & \textbf{55.13} & 53.69 & 58.08  \\
\hline
\end{tabular}
\end{table*}

\begin{table*}[t]
\caption{\label{table:results1c_f}{Teacher-1 A-Mistakes. Results under Without Collar and With Collar correspond to frame-based metrics, while the Event-Based section reports event-based metrics. Results showing performance of the rule-based model (RB), models using pitch contour features (CNN, CRNN (P), TCN (P)), and models using chromagram features (CRNN, TCN). Results for TCN$^{aug}$ show the effect of data augmentation on the TCN model.}}
\centering
\begin{tabular}{|c|c|ccc|ccc|ccc|ccc|}
\hline
 & \multicolumn{1}{c|}{\multirow{2}{*}{\textbf{Model}}} & \multicolumn{3}{c|}{\textbf{Scenario-1}} & \multicolumn{3}{c|}{\textbf{Scenario-2}} & \multicolumn{3}{c|}{\textbf{Scenario-3}} & \multicolumn{3}{c|}{\textbf{Scenario-4}} \\
\cline{3-14}
 &  & \textbf{F1} & \textbf{P} & \textbf{R} & \textbf{F1} & \textbf{P} & \textbf{R} & \textbf{F1} & \textbf{P} & \textbf{R} & \textbf{F1} & \textbf{P} & \textbf{R} \\
\hline
\multirow{5}{*}{\rotatebox[origin=c]{90}{\textbf{Without}}}
\multirow{5}{*}{\rotatebox[origin=c]{90}{\textbf{Collar}}}
& \textbf{RB}        & 54.85 & 56.22 & 53.55 &58.66 &41.72  &98.75  & 50.77 & 59.43 & 44.31 & 79.11 & 70.84 & 89.55 \\ 
& \textbf{CNN}     & 67.55 & 60.41 & 76.61 & 70.87 & 63.63 & 79.98 & 63.62 & 73.66 & 56.00 & 81.12 & 73.66 & 90.28 \\ 
& \textbf{CRNN}    & 65.58 & 58.03 & 75.38 & 71.11 & 63.94 & 80.10 & 64.03 & 61.93 & 66.24 & 82.16 & 73.27 & 93.50 \\ 
& \textbf{TCN}         & 66.37 & 57.66 & 78.19 & 71.51 & 63.59 & 81.68 & 62.62 & 63.49 & 61.76 & 81.66 & 73.54 & 91.79 \\ 
& \textbf{TCN$^{\text{aug}}$} & \textbf{77.09} & 66.20  & 92.28 & \textbf{80.86} & 70.46 &  94.88 & \textbf{76.83} & 72.02 &  82.34 & \textbf{91.58} & 86.53 &  97.26 \\
\hline
\multirow{5}{*}{\rotatebox[origin=c]{90}{\textbf{With}}}
\multirow{5}{*}{\rotatebox[origin=c]{90}{\textbf{Collar}}}
 & \textbf{RB} & 58.45 & 60.88 & 56.20 &66.46  &49.78  &100  & 57.63 & 69.70 & 49.12 & 82.30 & 74.74 & 91.56 \\
& \textbf{CNN}  & 73.18 & 65.19 & 83.41 & 75.92 & 68.89 & 84.56 & 72.58 & 83.69 & 64.07 & 85.76 & 77.03 & 96.72 \\ 
& \textbf{CRNN} & 72.03 & 63.28 & 83.59 & 76.87 & 69.54 & 85.91 & 75.09 & 74.15 & 76.06 & 85.79 & 76.42 & 97.79 \\ 
& \textbf{TCN}  & 71.88 & 62.48 & 84.60 & 77.23 & 69.10 & 87.54 & 73.02 & 74.17 & 71.91 & 86.03 & 76.85 & 97.72 \\ 
& \textbf{TCN$^{\text{aug}}$} & \textbf{80.80} & 71.35 & 93.14 & \textbf{84.76} & 75.62 & 96.40 & \textbf{81.88} & 77.89 & 86.30 & \textbf{94.69} & 90.76 & 98.97 \\
\hline
\multirow{5}{*}{\rotatebox[origin=c]{90}{\textbf{Event}}}
\multirow{5}{*}{\rotatebox[origin=c]{90}{\textbf{Based}}}
& \textbf{RB}     &30.77  &85.71  &18.75  &39.21&95.24  &24.69  &14.28  &60.00  &8.11  &34.37  &91.67  &21.15  \\
& \textbf{CNN}  & 54.57 & 42.94 & 74.84 & 56.01 & 46.01 & 72.79 & 41.34 & 28.28 & 77.14 & 76.20 & 68.03 & 86.82 \\
& \textbf{CRNN} & 57.32 & 48.57 & 70.36 & 63.64 & 57.28 & 71.65 & 57.26 & 47.86 & 71.43 & \textbf{86.67} & 86.36 & 87.72 \\
& \textbf{TCN}      & 64.45 & 59.14 & 70.86 & \textbf{67.25} & 61.72 & 73.95 & 57.72 & 47.86 & 72.86 & 85.90 & 85.90 & 85.90 \\
& \textbf{TCN$^{\text{aug}}$} & \textbf{65.59} & 66.04 & 71.11 & 60.66 & 65.51 & 60.34 & \textbf{63.54} & 59.48 & 75.30 & 82.33 & 85.12 & 82.39 \\
\hline
\end{tabular}
\end{table*}

\begin{table*}[t]
\caption{\label{table:results_event_t2}{Teacher-2 A-Mistakes}}
\centering
\begin{tabular}{|c|c|ccc|ccc|ccc|ccc|}
\hline
 & \multicolumn{1}{c|}{\multirow{2}{*}{\textbf{Model}}} & \multicolumn{3}{c|}{\textbf{Scenario-1}} & \multicolumn{3}{c|}{\textbf{Scenario-2}} & \multicolumn{3}{c|}{\textbf{Scenario-3}} & \multicolumn{3}{c|}{\textbf{Scenario-4}} \\
\cline{3-14}
 &  & \textbf{F1} & \textbf{P} & \textbf{R} & \textbf{F1} & \textbf{P} & \textbf{R} & \textbf{F1} & \textbf{P} & \textbf{R} & \textbf{F1} & \textbf{P} & \textbf{R} \\
\hline
\multirow{4}{*}{\rotatebox[origin=c]{90}{\textbf{Without}}}
\multirow{4}{*}{\rotatebox[origin=c]{90}{\textbf{Collar}}}
& \textbf{RB}        &27.45  &43.65  &20.03  &54.08  &64.17  &46.73  &51.70  &57.47  &46.98  &47.26  &52.61  &42.91  \\
& \textbf{CNN}     & 80.26 & 71.20 & 91.96 & \textbf{94.41} & 90.98 & 98.13 & 89.38 & 83.07 & 96.73 & 84.52 & 74.27 & 98.05 \\
& \textbf{CRNN}& 89.46 & 91.23 & 87.75 & 94.38 & 90.56 & 98.55 & 88.64 & 81.63 & 96.97 & 86.53 & 77.45 & 98.02 \\
& \textbf{TCN}     & \textbf{90.53} & 99.99 & 82.71 & 93.25 & 80.98 & 95.64 & \textbf{94.73} & 95.18 & 94.28 & \textbf{91.71} & 87.84 & 95.94 \\

\hline

\multirow{4}{*}{\rotatebox[origin=c]{90}{\textbf{With}}}
\multirow{4}{*}{\rotatebox[origin=c]{90}{\textbf{Collar}}}
 & \textbf{RB}     &27.71  &44.81  &20.05  &54.35  &64.76  &46.82  &51.84  &57.75  &47.03  &47.42  &52.92  &42.96  \\
& \textbf{CNN}  & 80.88 & 71.78 & 92.64 & \textbf{94.86} & 91.29 & 98.71 & 89.61 & 83.30 & 96.95 & 84.81 & 74.56 & 98.31 \\
& \textbf{CRNN} & 90.06 & 91.67 & 88.51 & 94.81 & 90.90 & 99.07 & 88.88 & 81.90 & 97.18 & 86.82 & 77.78 & 98.25 \\
& \textbf{TCN}  & \textbf{91.14} & 99.99 & 83.72 & 93.71 & 91.40 & 96.15 & \textbf{94.97} & 95.34 & 94.60 & \textbf{91.98} & 88.14 & 96.16 \\
 
\hline

\multirow{4}{*}{\rotatebox[origin=c]{90}{\textbf{Event}}}
\multirow{4}{*}{\rotatebox[origin=c]{90}{\textbf{Based}}}
& \textbf{RB}     &13.47 &10.45  &18.96  &14.18  &11.02  &19.87  &11.51 &9.11  &15.62  &13.63 &11.24  &17.32  \\

& \textbf{CNN}  &16.48  &11.13  &31.70  &16.66  &11.64  &29.32  &13.88  &9.53  &25.52  &13.91  &11.38  &17.91  \\
& \textbf{CRNN} & 45.51 & 39.51 & 53.65 &53.17  & 47.64 & 60.15 &50.16  & 45.46 & 55.94 &35.86 &29.91  &44.78  \\
& \textbf{TCN}      &\textbf{73.17}  &73.17  &73.17  &\textbf{75.97}  &73.87  &78.20  &\textbf{66.54}  &60.84  &73.43  &\textbf{69.50}  &67.67  &71.43  \\
\hline
\end{tabular}
\end{table*}


\section{Experiments}
Since the audio files are of variable length, we segment both the teacher's and the learner's audio into 4-second chunks. Prior to feature extraction, the teacher’s audio is tonic-normalized~\cite{param_PIMv1} using Eq.~\ref{eq: tonic_norm}, and the learner’s audio $ \mathbf{x}^{\ell}$ is truncated or zero-padded to match the duration of the corresponding teacher audio $ \mathbf{x}^{t}$. This ensures strict temporal alignment and equal dimensions for all teacher–learner audio pairs, which is essential for frame-level mistake detection. Then, we extract pitch and amplitude-based features as discussed in \ref{section: features}. The teacher and learner features are standardized to a zero mean and unit variance.  We split the training, validation, and test sets in a ratio of 70:15:15.

\subsection{Studies on Data Split Scenarios}
To assess generalization under varying conditions, we evaluate the models across four data-splitting scenarios, with experiments conducted independently for each teacher.
\begin{itemize}
    \item \textbf{Scenario 1} All learner recordings are pooled for respective teachers, and the recordings are randomly distributed into train and test sets. In this scenario, the recordings of same learner may appear in both train and test set. 
    \item \textbf{Scenario 2} Recordings from each learner appear exclusively in either training or testing, ensuring no overlap. This setup evaluates the model’s ability to generalize to entirely unseen learners.  
    \item \textbf{Scenario 3} For each learner, their recordings are ordered chronologically. Early (simpler) lessons are used for training, and later (advanced) lessons for testing. This examines whether a model trained on simpler material can detect mistakes in more complex contexts.  
    \item \textbf{Scenario 4} The reverse of Scenario 3. Advanced lessons are used for training, and beginner lessons for testing. This investigates whether training on more complex lessons enables the model to perform better on simpler ones.  
\end{itemize}
The Train-Test distribution of ground truth for all 4 scenarios is illustrated in Figures \ref{fig:scenario_dist_001}  and \ref{fig:scenario_dist_002}.

\subsection{Cross-teacher Studies} \label{ssec:crossteacherstudies}
To assess cross-teacher generalization, we train on data from one teacher (in-teacher) and test on the other teacher’s data (out-teacher), following Scenario~1,
thereby evaluating robustness to differences in teaching style, annotation practices, and labeling criteria.

\begin{table}[ht]
\centering
\caption{Cross-teacher evaluation results for frame-based and event-based metrics.}
\renewcommand{\arraystretch}{1.3}
\resizebox{8.8cm}{!}{%
\begin{tabular}{|c|ccc|ccc|}
\hline
\multirow{2}{*}{\textbf{Train $\rightarrow$ Test}} & \multicolumn{3}{c|}{\textbf{Frame-based}} & \multicolumn{3}{c|}{\textbf{Event-based}} \\ \cline{2-7}
& \textbf{F1} & \textbf{P} & \textbf{R} & \textbf{F1} & \textbf{P} & \textbf{R} \\ \hline
Teacher 1 $\rightarrow$ Teacher 2 & 71.83 & 75.80 & 78.69 & 56.64 & 56.48 & 57.17 \\
Teacher 2 $\rightarrow$ Teacher 1 & 70.00 & 67.34 & 88.43 & 40.06 & 40.34 & 39.92 \\ \hline
\end{tabular}
}
\label{tab:cross_teacher}
\end{table}

\begin{figure*}[t]
\centering
\includegraphics[width=\textwidth]{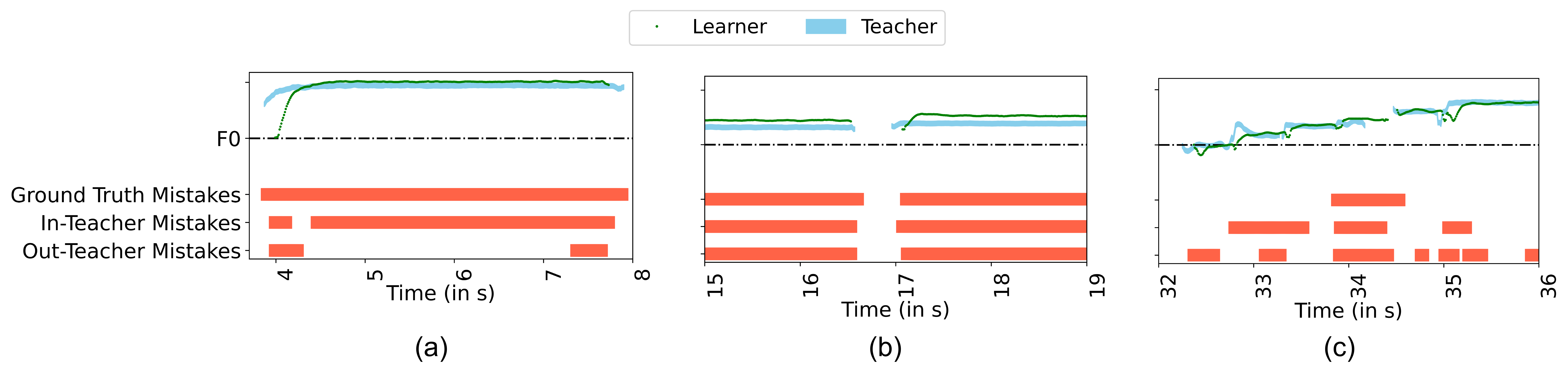}
\caption{\label{fig: mistakes}{Samples illustrating cross-teacher analysis: (a) cross-teacher differences, (b) cross-teacher similarities, (c) a difficult lesson. Here, in-teacher and out-teacher mistakes refer to the outputs of models trained on the same (as ground truth) and different teachers, respectively.}}
\end{figure*}

\section{Results and Discussion}
We observe that across teachers and split scenarios, learning-based models consistently outperform the RB baseline on frame-based metrics
(Tables~\ref{table:results_frame_f_t1}--\ref{table:results_event_t2}). This gap is expected as RB relies on fixed thresholds applied frame-by-frame and lacks temporal context, making it sensitive to local pitch-tracker noise and micro-variations. In contrast, neural network-based models leverage context and learn data-driven decision boundaries.

Among learned models, TCN achieves the strongest and most stable performance in most settings, attributed to its dilated convolutions that capture long-range temporal dependencies and better enforce event continuity. CRNN variants typically rank next, followed by CNN. Event-based scores also follow this pattern but are generally lower for RB. This is because it has no notion of temporal hysteresis or continuity, leading to over-segmentation into many short fragments.

We observe in Tables~\ref{table:results_frame_f_t1}--\ref{table:results_event_t2} that for frequency (F) errors, models using pitch contours generally outperform those using chroma features. This is because monophonic conditions and reliable tonic normalization enable precise $f_0$ tracking. Moreover, pitch contours preserve fine-grained intonation deviations, which get smeared in chromagram representation. As a result, pitch contours improve discriminability in borderline cases. In addition, octave-invariant normalization prevents penalizing octave shifts that are not considered mistakes pedagogically.

We also note that Teacher~2 models tend to achieve higher F1 in several settings, which correlates with larger data volume for Teacher~2. More training data improves coverage of mistake patterns and regularizes model estimates. Finally, hysteresis thresholding improves event-based metrics by suppressing spurious short activations and stabilizing segment continuity.

\subsection{Effect of Collar, Post-processing, and Data Augmentation}
Introducing a collar improves frame-wise metrics across models (Tables~\ref{table:results_frame_f_t1}--\ref{table:results_event_t2}), reflecting realistic tolerance for annotation boundary uncertainty.
We use a collar $T_c=80$ ms for frame-based and $200$ ms for event-based metrics to reflect typical annotation imprecision at the frame and event levels, respectively. 
In addition, Table~\ref{table:results1c_f} highlights the effectiveness of data augmentation: the model trained with augmented data ($\mathrm{TCN}^{aug}$) consistently outperforms its non-augmented counterpart, demonstrating that perturbing static notes reduces class imbalance. It is important to note that augmentation was applied only for Teacher-1, since a substantial class imbalance was observed in this case, while Teacher-2 did not exhibit such imbalance.

\subsection{Analysis of Data Split Scenarios}
We evaluate four data splits within each teacher’s corpus:

\textbf{Scenario~1 vs. Scenario~2}: Deep models show comparable performance across these settings, indicating that they do not overfit to specific learner characteristics and can generalize to unseen learners. This suggests that the models capture mistake patterns that are shared across learners rather than idiosyncratic voice traits.

    \textbf{Scenario~3 vs. Scenario~4}: Models generally perform better in Scenario~4, especially for frequency mistakes. Training on complex lessons exposes models to a wider range of melodic contexts and more diverse mistake manifestations, improving generalization to simpler lessons. This asymmetry is musically plausible: expertise developed on complex material readily covers fundamentals, whereas the reverse transfer is weaker.

\subsection{Cross-Teacher Studies:}
Cross-teacher transfer (train on Teacher~1, test on Teacher~2, and vice versa; Table~\ref{tab:cross_teacher}) shows a clear drop in frame-based F1 compared to in-teacher results, while event-based F1 is less affected. This difference reflects teacher-specific tolerance policies: Teacher~2 applies stricter pitch (12 cents) and amplitude (6 dB) thresholds, whereas Teacher~1 allows larger deviations (22 cents, 30 dB) (Table~\ref{table:baseline_hyperparam}). As a result, Teacher~2 often flags brief or subtle deviations, while Teacher~1 emphasizes more sustained errors. These differences, especially around rapid pitch motion, phrase boundaries, and ornamented renditions, increase annotation ambiguity and aleatoric uncertainty.  
Case studies in Fig.~\ref{fig: mistakes} illustrate three patterns: (a) out-teacher models (Train: Teacher~1, Test: Teacher~2) are more lenient near boundary deviations than in-teacher (Train and test on Teacher~2) models; (b) both models agree on clear mistakes; (c) both struggle on dense, ornamented passages. Visualizing detected frequency errors on pitch contours provides interpretable, teacher-aligned feedback, which is pedagogically useful.
Ultimately, these cross-teacher variations suggest that the definition of an ``error" is inherently teacher-specific; the performance gap observed across instructors reflects divergent evaluative tolerances and pedagogical priorities rather than being a mere consequence of model limitations.

\begin{table}[t]

\caption{\label{table:baseline_hyperparam}
Teacher-specific optimal thresholds for the rule-based baseline. 
$\beta$ is an octave-normalized pitch tolerance (shown also in cents); 
$\gamma$ is an amplitude tolerance (shown also in \textnormal{dB}).
}
\centering
\scriptsize
\setlength{\tabcolsep}{5pt} 
\renewcommand{\arraystretch}{1.2} 
\begin{tabular}{|c|c|c|c|c|c|}
\hline
\textbf{Teacher} &        & \textbf{Scenario-1} & \textbf{Scenario-2} & \textbf{Scenario-3} & \textbf{Scenario-4} \\ \hline
\multirow{2}{*}{1} 
  & $\beta$ (cents)  & 0.018 (21.6) & 0.01 (12.0) & 0.017 (20.4) & 0.01 (12.0) \\ 
  & $\gamma$ (dB)    & 4 (24.1)     & 5 (30.1)    & 5 (30.1)     & 6 (36.1)    \\ \hline
\multirow{2}{*}{2} 
  & $\beta$ (cents)  & 0.01 (12.0)  & 0.01 (12.0) & 0.01 (12.0)  & 0.01 (12.0) \\ 
  & $\gamma$ (dB)    & 1 (6.0)      & 1 (6.0)     & 1 (6.0)      & 1 (6.0)     \\ \hline
\end{tabular}
\end{table}


\section{Conclusion and Future Work}

In this work, we present a framework for automatic detection of singing mistakes in the context of Indian Art Music pedagogy, supported by the curated M3 dataset of synchronized teacher–learner recordings. By benchmarking rule-based and deep learning approaches, we demonstrate the superiority of TCN in capturing pitch and amplitude errors, while proposing a collar-based evaluation strategy that accounts for annotation uncertainties. A systematic analysis of learner errors and cross-teacher studies further highlights the pedagogical value of the system in offering interpretable and context-aware feedback

Future research will focus on expanding the framework to accommodate learner improvisation and stylistic variations beyond strict imitation, as well as extending the methodology to other instruments and vocal traditions in Indian classical music.
Addressing timing-related mistakes will require the development of beat or tāla-aware modeling approaches.

Longitudinal classroom studies will be critical to assessing the practical benefits of automated feedback in real teaching–learning contexts.
Finally, integrating explainable models and real-time feedback mechanisms will ensure that such systems remain pedagogically effective, complementing rather than replacing the essential human elements of music education.

\section*{Acknowledgments}
We would like to thank Kajal Nagra, Renu Chavan, and Shivnarayan Pandey for their assistance in dataset preparation.

\bibliographystyle{IEEEtran}
\bibliography{references}

\end{document}